\documentstyle[prl,aps]{revtex}
\input{psfig.sty}
\draft
\begin{document}
\twocolumn[\hsize\textwidth\columnwidth\hsize\csname@twocolumnfalse\endcsname

\title{Formation of longitudinal structures in  granular flows}
\author{Igor S. Aranson$^1$  and Lev S. Tsimring$^2$}
\address{
$^1$ Argonne National Laboratory,
9700 South  Cass Avenue, Argonne, IL 60439 \\
$^2$ Institute for Nonlinear Science, University of California,  
San Diego, La Jolla, CA 92093-0402 }
\date{\today}

\maketitle

\begin{abstract} 
In the framework of the theory of partially fluidized granular flows we
study the formation of longitudinal structures observed
experimentally by Forterre and Pouliquen in a flow down a
rough inclined plane.  We show that the formation of longitudinal
structures is related to the positive feedback between the fluidization
rate and the lateral stress (side pressure), which leads to a convective
instability.  Our theory explains main experimental features, such as
appearance and amplification of the structure at some distance from the
outlet  and non-stationary behavior of the structures.
\end{abstract}


\narrowtext
\vskip1pc]

The dynamics of granular media has been an active  area of research
for physicists \cite{jnb} and engineers
\cite{nedderman}.  One of the most interesting phenomena pertinent to
many granular systems is the  transition from a static equilibrium to a
granular flow.  There has been a number of experimental studies of flows
in large sandpiles\cite{bagnold,radj} as well as in thin layers of
grains on inclined surfaces \cite{daerr,daerr1,pouliquen,forterre}.  
Recent experiments with granular flows on a rough inclined plane revealed
a new striking phenomenon: formation of non-stationary longitudinal
structures at some distance downstream from an outlet
\cite{forterre}.  Authors of \cite{forterre} proposed an explanation 
of the instability mechanism based on analogy with thermal convection 
in fluids:  a rapid shear granular flow 
leads to the increase of the granular temperature 
near the rough bottom. Because of the intrinsic dissipative
nature of collisions between particles, the granular temperature should decay
away from the bottom, creating necessary conditions for the ``thermal granular
convection'' (cf. \cite{wildman,meerson}).  Using hydrodynamic
equations obtained  from the kinetic theory for dilute granular gases,
this instability was studied analytically and numerically \cite {forterre1}.
Although shear flow activated thermal granular convection could be 
a useful concept for the interpretation of  experimental results, 
the theory \cite{forterre1} did not address  two important 
observations: (1) experiments as well as simulations \cite{ertas} 
show that fluctuations of velocity are more
significant near the free surface of the granular flow, and so the
granular temperature may in fact be higher at the top rather than at the
bottom, (2) longitudinal structures 
appear at some distance downstream from the outlet
and exhibit complex spatio-temporal dynamics.

In this Letter we demonstrate that the observed transverse 
instability  can be described within the framework of the continuum
theory of partially fluidized flows proposed by us in Refs.\cite{at1,at2}
without invoking the concept of granular temperature more appropriate
for dilute gas-like granular flows.
In this theory, we introduce an equation for the order
parameter which characterizes the phase state of the granular matter. 
In a certain range of parameters, the shear flow described by these 
equations, is unstable with respect to transverse perturbations. We
show that this instability is of convective nature, i.e. small
perturbations grow downstream while remaining small in the laboratory
frame. Thus, the ``rolls'' appear at some finite distance from the
outlet.  This distance depends on the noise level
and flow conditions. Since the pattern structure is
determined by random fluctuations near the outlet, the resulting pattern 
is always non-stationary in the laboratory frame, very similar to 
that observed experimentally. 


{\it Model}. 
The starting point of our theory is the momentum conservation equation
\begin{equation}
\rho_0 D v_i/Dt=\frac{\partial
\sigma_{ij} }{\partial x_j}+ \rho_0  g_i, \;\;j=1,2,3.
\label{elastic}
\end{equation}
where $v_i$ are the components of velocity, $\rho_0=const$ is the
density of material (in the following we set $\rho_0=1$), 
${\bf g}$ is acceleration of gravity, and 
$D/Dt=\partial_t+v_i\partial_{x_i}$ denotes the material
derivative. Since the relative density fluctuations are small, the
velocity obeys the incompressibility condition $\nabla\cdot {\bf v}=0$. 
The stress tensor is represented as a sum of the hydrodynamic part proportional
to the flow strain rate and the strain-independent  part,
\begin{equation}
\sigma_{ij}=
\eta \left(\frac{\partial v_i }{\partial x_j}  +
\frac{\partial v_j }{\partial x_i} \right) + \sigma_{ij}^s,
\label{sigma}
\end{equation}
where $\eta$ has the meaning of the viscosity coefficient.  
At this point we introduce the order parameter (OP) $\rho$ which describes the
``phase state'' of the granular matter: it varies from 0 in the
``liquid'' phase to 1 in the ``solid'' phase. 
We interpret the OP as the relative
local density of static contacts among the grains.  
In the solid state ($\rho=1$) the strain-independent part should
coincide with the ``true'' static stress tensor $\sigma_{ij}^0$ for the
immobile grain configuration in the same geometry, and in the completely
fluidized state ($\rho=0$) we should have $\sigma^s_{ij}=-\Pi\delta_{ij}$
($\Pi$ is the hydrodynamic pressure). According to our assumption, 
the off-diagonal elements of the strain-independent part of the stress 
tensor obey 
the conditions 
$\sigma_{ij}^s = \rho \sigma_{ij}^0$ for $i \ne j$. In the solid 
state the normal stresses $\sigma_{ii}^0 $ do not in general coincide.
For the weakly-fluidized state ($\rho \to 1$) the normal stresses 
are close to the static values, however some dependence on the order 
parameter $\rho$ (i.e. degree of fluidization)  may appear. 
In the  first order in $1-\rho$ one  writes for the diagonal elements of
the stress tensor
\begin{equation} 
\sigma_{ii}^s = \sigma_{ii}^0 
+\alpha_i(1-\rho) + O((1-\rho)^2)
\label{stress1} 
\end{equation} 
where $\alpha_i$ 
characterizes the response of the normal stresses 
on fluidization. Since fluidization is accompanied by the 
decrease in the number of static contacts among granules (i.e. 
dilution), the normal stresses should decrease with fluidization, i.e.
$\alpha_i
>0$. It also agrees with the observation in Ref. \cite{forterre} that 
the crests of the surface deformations correspond to a more dilute 
granular state.

According to Ref.\cite{at1,at2}, the equation for the order parameter 
$\rho$ is taken in the form 
\begin{equation}
\dot{\rho}+ v \nabla \rho 
=\nabla^2\rho+\rho(1-\rho)(\rho-\delta),
\label{op-eq1}
\end{equation}
where $\delta=(\phi-\phi_0)/(\phi_1-\phi_0)$ is the control parameter,
$\phi=\max|\sigma_{mn}^0/\sigma_{nn}^0|$ is the maximum ratio of shear
to normal stresses in  the bulk,  and $\tan^{-1}\phi_{1,2}$ are static
and dynamic repose angles characterizing the granular material.
Parameter $\phi$ which enters  the Mohr-Coloumb yield
condition\cite{nedderman}, in the context of our theory is equivalent to
a melting temperature in the theory of phase transition.

\begin{figure}[h]
\centerline{ \psfig{figure=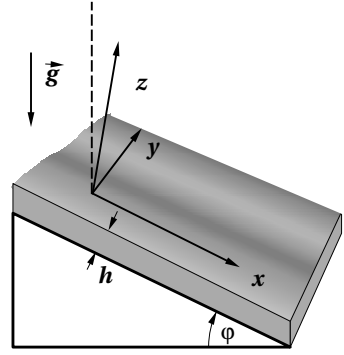,height=2.5in}}
\caption{Schematic representation of a chute geometry. 
$z$-axis is normal to the chute bottom, dashed line is 
parallel to the direction of gravity. 
}
\label{setup}
\end{figure}

Following the analysis 
of Ref. \cite{at1},  we consider a layer of dry cohesionless grains
on an inclined rough surface (see Fig.\ref{setup}). However, now we 
assume that the layer thickness $h$ can vary in both 
$x$ and $y$ directions.  The momentum conservation equation
(\ref{elastic}) in
the stationary regime yields the force balance conditions
\begin{eqnarray}
\label{eqv}
\sigma_{xz,x}+\sigma_{yz,y}+\sigma_{zz,z} &=& g \cos \varphi,  \\
\sigma_{xx,x}+\sigma_{xy,y} + \sigma_{xz,z} &=& 
-g \sin \varphi,   \\
\sigma_{xy,x}+ \sigma_{yy,y} + \sigma_{yz,z} &=& 0,
\label{eqv1}
\end{eqnarray}
where the subscripts after commas mean partial derivatives, 
and $z=0$ corresponds to the bottom of the layer. 
We assume that  variations of the 
layer thickness along the $y$ direction are small. 
Accordingly, there will be 
a small component of velocity along $y$ direction 
and corresponding stress  components $\sigma_{yz}$ and 
$ \sigma_{yy}$. 
In the avalanche problem considered in Ref. \cite{at1} these terms were 
irrelevant. We show later that the transverse flux is crucial 
for the explanation of longitudinal structures. 

In the first order  in $h_x,h_y$
Eqs. (\ref{eqv})-(\ref{eqv1})  yield 
\begin{eqnarray} 
\sigma_{zz} & \approx &  g \cos \varphi (z-h) \;,\;
\sigma_{xz}  \approx  -g \sin \varphi (z-h) , \nonumber \\
\sigma_{yz}  & =&   
 \int_{z}^h dz \sigma_{yy,y}.
\label{sigma2} 
\end{eqnarray} 
Here we used the conditions $\sigma_{xx}=const, \sigma_{xy}=const$ and
 $\sigma_{zz}=\sigma_{xz}= \sigma_{yz}=0$ on free surface $z=h$. 

For the
chute geometry Fig. \ref{setup} parameter $\delta$ in Eq. (\ref{op-eq1}) can be specified. 
For $h=const$ 
there is a simple relation between shear and
normal stresses, $\sigma_{xz} = -  \tan \varphi \sigma_{zz}$.
The most ``unstable'' yield
direction is parallel to the inclined plane,
i.e. $\phi=|\sigma_{xz}/\sigma_{zz}|$ and $\delta=\delta_0=
(\tan \varphi-\phi_0)/(\phi_1-\phi_0)$. 
If $h$ is  a  slowly varying function of $x$ and $y$, one 
obtains for the parameter $\delta$: 
\begin{equation} 
\delta=\delta_0- \beta h_x + O(h_x^2,h_y^2) 
\label{delta1}
\end{equation} 
where $\beta=1/(\phi_1-\phi_0)$ (see \cite{at1,at2} for more details). 

{\it Thin layer solutions} of  Eq. (\ref{op-eq1}) for the chute
geometry Fig. \ref{setup} are subject to the following boundary
conditions (BC):  no-flux condition $\rho_z= 0$ at the free surface $z=h$,
and  no-slip condition $\rho=1$ at the bottom of the chute $z=0$
(a granular medium is
assumed to be in a solid phase near the rough surface).  
The velocity profiles corresponding to a stationary profile of $\rho(z)$,
can be easily found from Eq. (\ref{sigma}),(\ref{stress1}),
\begin{eqnarray} 
\eta \frac{\partial v_x}{\partial z} &=&  -g \sin \varphi (z-h) - \rho\sigma
_{xz}^0= -g \sin \varphi (1-\rho)(z-h), \nonumber \\
\eta \frac{\partial v_y}{\partial z} &=& - 
\int_{z}^h \alpha_y \partial_y
\rho dz 
\label{vel}
\end{eqnarray} 
In the derivation of Eq. (\ref{vel})  we used $\sigma_{yz}^s=0$ for a
flat layer. 
The components of the flux ${\bf J} =
\int _{0}^h {\bf v}(z) dz$ 
are   
\begin{eqnarray}
J_x &=& 
-\frac{g \sin \varphi }{\eta} \int ^{h}_0 dz \int _{0}^z
(1-\rho(z^\prime)) (z^\prime-h) d z^\prime \nonumber  \\
J_y &=&   
-\frac{\alpha_y}{\eta}\int ^{h}_0 dz   \int _{0}^z 
dz^{\prime \prime}  \int_{z^{\prime \prime}}^h \partial_y  
\rho dz^\prime 
 \label{j}
\end{eqnarray}

For thin weakly-fluidized layers
(see \cite{at1,at2}),  we can look for solution of
Eq.(\ref{op-eq1}) in the form 
\begin{equation} 
\rho =1 - A \sin\left(\frac{\pi}{2 h} z\right)+ \mbox{h.o.t.},
\label{form1} 
\end{equation} 
where $A\ll 1$ is a slowly varying function of $t,\ x$, and $y$. 
Substituting ansatz  (\ref{form1}) into Eq. (\ref{op-eq1}) and applying
orthogonality conditions \cite{at1,at2}), we obtain 
\begin{equation} 
A_t = \Lambda A+ \nabla^2_\perp A
+\frac{8(2-\delta)  }{3 \pi} A^2 -\frac{3 }{4}  A^3 
- \bar \mu
h^2 A \partial_x A
\label{A1} 
\end{equation} 
where $\nabla^2_\perp = \partial_x^2+\partial_y^2$, $\Lambda  =
\delta-1 - \frac{\pi^2}{4 h^2}$, 
$\bar \mu=
 (3 \pi^2-16)g\sin\varphi/3 \pi^3\eta=0.146g\sin\varphi/\eta $.  
The mass conservation yields 
\begin{equation} 
\frac{\partial h}{\partial t} = 
-\nabla \cdot {\bf  J}= - \partial_x J_x  -\partial_y J_y  
\label{conser}
\end{equation} 
where the flux  components $J_{x,y}$ calculated from Eq. (\ref{j}) in the thin 
layer approximation, are  
\begin{eqnarray} 
J_x= \mu  h^3 A\;,  
J_y= -\alpha_1 h^2 A h_y + \alpha_2 h^3  A_y 
\label{j2}
\end{eqnarray}
$\mu= 
2(\pi^2-8)g\sin\varphi /\eta \pi^3\approx
0.12 g\sin\varphi /\eta$, $\alpha_1= \alpha_y (\pi^3+8 \pi-48)/2 \pi^3 \eta
\approx 0.13
\alpha_y/\eta $, $\alpha_2= 8\alpha_y /  \pi^3 \eta\approx 0.26 \alpha_y/\eta$.

\begin{figure}[h]
\centerline{ \psfig{figure=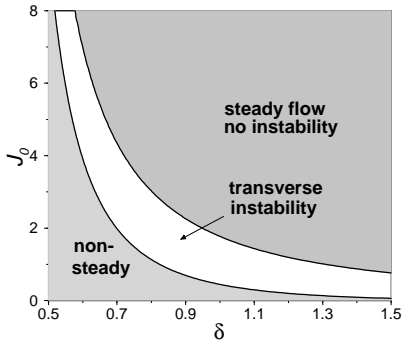,height=2.in}}
\caption{
Phase diagram in $\delta - J_0$ plane. 
}
\label{phase}
\end{figure}

 Our model has two control
parameters, the downhill mass flux $J_0$ at the outlet $x=0$, and $\delta$. 
For large enough $J_0$ and $\delta$, Eqs. (\ref{A1}),(\ref{conser}) possess a
steady-state solution $A=A_0, h=h_0$ (see Fig.\ref{phase}). The
stability of this solution can be examined by
substituting the ansatz $\{A,h\} = \{A_0,h_0\}+ \{\zeta,\xi\} \exp (\lambda t + 
i k_x x +i k_y y)$,
where $\{\zeta,\xi\}$ is a small perturbation. For long-wave
perturbations ($k_x,k_y\ll 1$) in the leading order one 
obtains for the growth rate $\lambda$ 
\begin{equation} 
\lambda = ik_x \mu h_0^2A_0+  B k_y^2 + ...
\label{lambda1} 
\end{equation} 
where 
$B=\pi^2  \alpha_2 [16(2-\delta) /3 \pi  -3 A_0]^{-1} - \alpha_1  h_0^2
A_0$.
As seen from Eq. (\ref{lambda1}), 
pure longitudinal perturbations $k_y=0$ are neutrally
stable, but for $B>0$ the perturbations with transverse component of the
wave number can be unstable.  For pure transverse perturbations
($k_x=0$) the instability is aperiodic ($\lambda$ is real), as in
thermo-convection in ordinary fluids.  The physical meaning of the
instability is the following: decrease of $\rho$ (increase of
fluidization) corresponds to the decrease of the lateral pressure
($-\sigma_{yy}$) which  in turns lead to the increase of the height $h$
and further fluidization.  The phase diagram of the instability on
$\delta-J_0$ plane is shown in Fig. \ref{phase}.  The transverse
instability exists in the band  restricted from above by the condition
$B=0$ and from below by the existence of steady-state solution
\cite{comment}. 

Complexity of $\lambda$ and asymmetry of the problem in 
$x$-direction  signal the possibility of the convective nature of the 
instability: the perturbations may
grow in the moving frame and decay in the laboratory frame
(see, e.g. \cite{landau,akw}). 
As a test for convective instability one has to examine the value of
$\lambda$ for the saddle point of the function $\lambda(k_x,k_y)$
in the complex $k_x$ plane, i.e.
$\partial \lambda /\partial k_x=0$. 
In fact, it is easy to see that if the instability is weak (for
$\alpha_y\ll 1$) and the downhill flux $J_0$ is finite, the instability has to
be convective. 
In the presence of persistent fluctuations, e.g. at the flow outlet, 
the perturbation will be spatially amplified  down the flow. 

Let us  estimate the spatial growth rate.  Consider the
unstable perturbations caused by a 
small stationary $y$-periodic force with wavenumber 
$k_0$ localized at $x=0$.  
Calculations show that 
the linearized solution $\zeta$ will be described by the 
integral 
\begin{equation}
\zeta \sim \int _{-\infty }^\infty \exp[i{\bf k r}]
\frac{dk_x}{ \lambda({\bf k} )}
\left( 1-\exp[ \lambda({\bf k} )t ] \right)      
\label{integral}
\end{equation} 
The non-stationary part $\sim \exp[i\lambda({\bf k})t]$ decays in time 
due to the convective character of the instability, and the 
integral yields
\begin{equation}
\zeta \sim   \exp [ i k_0 y + i k^* x] 
\label{integral1}
\end{equation} 
where $k^*$ is found from $\lambda (k_x=k^*, k_0) = 0 $. It is easy to
see from Eq.(\ref{lambda1}) that,  for small $k^0$, $k^* \approx - i B k_0^2/\mu
h_0^2A_0$.  Since $k^*$ is imaginary, perturbations from a stationary
source at $x=0$  grow exponentially downhill.  For random perturbations
introduced at the outlet the typical scale of the pattern  will be
determined by the most unstable wavenumber  in the transverse direction.
However, the pattern will remain non-stationary due to intrinsically
random nature of the noise. 

\begin{figure}[h]
\centerline{ \psfig{figure=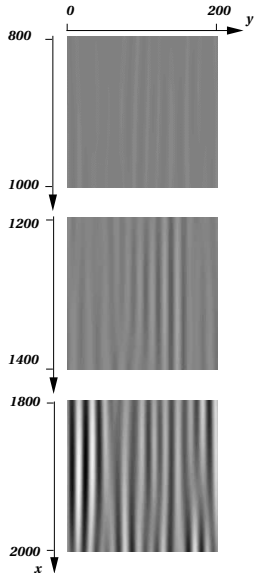,height=3.in}}
\caption{
Grey-scale snapshots of $h(x,y)$ (white corresponds to larger $h$), 
for $\delta=1.5$, $\mu=0.025 $, $\beta=3.14 $, $J_0=1.75$,
$\alpha_y=0.04$. Equilibrium value of 
the layer thickness $h_0 \approx 4.45$. 
}
\label{stab}
\end{figure}

\begin{figure}[h]
\centerline{ \psfig{figure=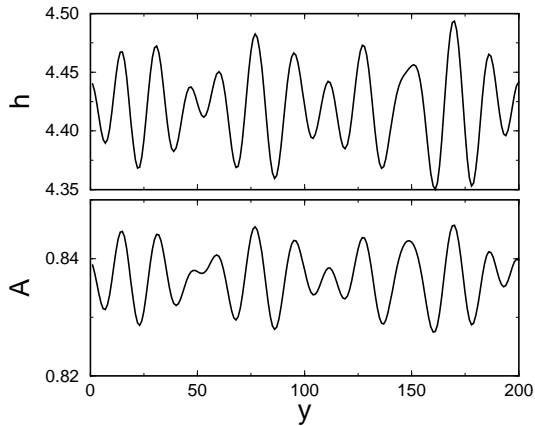,height=2.5in}}
\caption{
Profiles of $h(y)$ and $A(y)$ at $x=2000$. 
}
\label{Fig4}
\end{figure}

We studied Eqs. (\ref{A1}),(\ref{conser}) numerically. The simulations
were performed in a large system, more than $2000$ dimensionless units in
$x$-direction (downhill), and $200$ units in $y$-direction. 
Fixed flux boundary conditions were imposed at the outlet at $x=0$. 
Selected results are presented in Figs. \ref{stab},\ref{Fig4}. 
Figure \ref{stab} illustrates spatial amplification of perturbations 
downstream and formation of longitudinal structures. As it follows 
from our simulations,  far away from the outlet the structures remain 
non-stationary and exhibit spatio-temporal dynamics which are 
very similar to observed in Ref. \cite{forterre}. The  
profiles of $h$ and $A$ vs $y$ are shown in Fig. \ref{Fig4}. In
agreement with experiment, crests in $h$ corresponds 
to crests in $A$, i.e. to more fluidized regions of flow.

We demonstrated that formation of longitudinal 
structures in granular flow down a rough inclined plane is the result of 
noise amplification and saturation  due to the convective instability. 
The mechanism of the instability is related to the dependence of the local
pressure on the fluidization rate controlled by the order parameter. 
We conjectured a simple linear relation between these quantities.
It would be of interest to verify this relation in physical experiment
and molecular dynamic simulations.
The convective character of the  instability suggests that the flow 
can be controlled by small systematic perturbations introduced at the 
outlet.

In recent paper \cite{shen}, a novel fingering instability in a thin
granular layer inside a horizontal rotating cylinder was reported.
We believe that the nature of this instability fundamentally is the same
as described here. The only important difference is that due to backflow 
the perturbation can propagate upstream, and thereby the convective
instability becomes ``global''. This may explain why this fingering
instability occurs in a ``short'' system as compared with inclined layer
experiment \cite{forterre}.

This research was supported by the Office of the
Basic Energy Sciences at the US DOE, grants W-31-109-ENG-38 and
DE-FG03-95ER14516.
Simulations were performed at the National Energy Research 
Scientific Computing Center. 

\end{document}